\documentclass[review, 12pt]{myarticle}

\usepackage{amssymb}
\usepackage{bm}
\usepackage[colorlinks=true, linkcolor=red, anchorcolor=blue,citecolor=green]{hyperref}
\usepackage{amsmath}
\usepackage{subcaption}

\usepackage{cleveref}
\usepackage{multirow}
\usepackage{color,soul}
\usepackage[table]{xcolor}

\usepackage{stmaryrd}

\newcommand{\poly}[2]{\mathcal{P}_{#1}^{#2}}
\newcommand{\tria}[2]{\mathcal{T}_{#1}^{#2}}

\newcommand{\pedge}[1]{\mathcal{E}_{#1}}

\usepackage[switch]{}

\newcommand\E{{\bf e}}

\newcommand\M{{\bf m}}
\newcommand\N{{\bf n}}

\newcommand\T{{\bf t}}

\newcommand\X{{\bf x}}

\newcommand\CC{{\bf C}}

\newcommand\FF{{\bf F}}

\newcommand\GG{{\bf G}}

\newcommand\be{\begin{equation}}
\newcommand\nd{\end{equation}}
\newcommand\bed{\begin{displaymath}}
\newcommand\ndd{\end{displaymath}}

\newcommand\ba{\begin{array}}
\newcommand\ea{\end{array}}
\newcommand\bea{\begin{eqnarray}}
\newcommand\nda{\end{eqnarray}}

\usepackage{algorithm2e}
\RestyleAlgo{ruled}
\LinesNumbered
\SetKwComment{Comment}{/* }{ */}
\SetKw{Continue}{continue}
\SetKw{Break}{break}
\SetKw{And}{and}
\SetKw{Or}{or}

\begin{document}
% \captionsetup[figure]{labelfont={bf},name={Fig.},labelsep=period}
% \captionsetup[table]{labelfont={bf},name={Table.},labelsep=period}
\sethlcolor{yellow}
\setstcolor{red}
\soulregister\cite7 % for \cite
\soulregister\ref7
%\UseRawInputEncoding

\begin{frontmatter}

\title{Exact computation of the color function for triangular element interfaces}

\author[a]{Jieyun Pan\corref{cor1}}
\ead{yi.pan@sorbonne-universite.fr}

\author[a]{D\'{e}sir-Andr\'{e} Koffi Bi}

\author[a]{Ahmed Basil Kottilingal}

\author[a]{Serena Costanzo}

\author[c]{Jiacai Lu}
%\ead{jiacai.lu@jhu.edu}

\author[d]{Yue Ling}
%\ead{stanley\_ ling@sc.edu}

\author[e]{Ruben Scardovelli}
%\ead{ruben.scardovelli@unibo.it}

\author[c]{Gr\'{e}tar Tryggvason}
%\ead{gtryggv1@jhu.edu}

\author[a,b]{St\'{e}phane Zaleski}
%\ead{stephane.zaleski@sorbonne-universite.fr}

\cortext[cor1]{Corresponding author}
\address[a]{Sorbonne Universit\'{e} and CNRS, Institut Jean Le Rond d'Alembert UMR 7190, F-75005 Paris, France}
\address[b]{Institut Universitaire de France, Paris, France}
\address[c]{Department of Mechanical Engineering, Johns Hopkins University, Baltimore, MD, USA}
\address[d]{Department of Mechanical Engineering, University of South Carolina, Columbia, SC, USA}
\address[e]{DIN - Lab. di Montecuccolino, Universit\`a di Bologna, I-40136 Bologna, Italy}

\begin{abstract}
The calculation of the volume enclosed by curved surfaces discretized into triangular elements, and a cube is of great importance in different domains, such as computer graphics and multiphase flow simulations. We propose a robust algorithm, the Front2VOF (F2V) algorithm, to address this problem. 
The F2V algorithm consists of two main steps. First, it identifies the polygons within the cube by segmenting the triangular elements on the surface, retaining only the portions inside the cube boundaries.
Second, it computes the volume enclosed by these polygons in combination with the cube faces.
To validate the algorithm's accuracy and robustness, we tested it using a range of synthetic configurations with known analytical solutions.
\end{abstract}


\end{frontmatter}

% \linenumbers

% main text
\section{Introduction}

Calculating the volume enclosed by curved surfaces and a cube in numerical computations is of great interest in many different domains, such as multiphase flows and computer graphics.
In multiphase flow simulations based on the one-fluid model, physical properties are determined by the color function, representing the volume fraction of the reference phase within each cell. In Volume-of-Fluid (VOF) methods, the color function is directly evolved based on the governing equation. In contrast, in the Front-Tracking methods, the color function needs to be computed based on the positions of Lagrangian elements on the interface.

In the original Front-Tracking method proposed by Unverdi and Tryggvason \cite{Unverdi_1992_100}, which has been implemented in our free software \texttt{PARIS} simulator \cite{Aniszewski_2021_263}, the color function is obtained by solving a Poisson equation:
% ----------
\begin{gather}
\nabla^2 I = \nabla \cdot \GG_f,\\
\GG_f = (\nabla I)_f,
\end{gather}
% ----------
where $I$ is the color function, and $\GG_f$ represents the gradient of the color function on the background Eulerian grid, obtained by interpolating its value from the front. Since this gradient is directly computed at the front and then distributed to the fixed grid, the color function becomes smeared. Additionally, solving a Poisson equation, even within a narrow band domain near the interface, is time-consuming. 

Instead of calculating the color function by solving a Poisson equation, it is possible to leverage geometric information of the interface, such as the positions of marker points and the normal vectors of Lagrangian elements, to directly construct the color function. For two-dimensional problems where the interface is represented by line segments, this approach has been well addressed in prior work \cite{Tryggvason_2011_book, Aulisa_2003_188, Aulisa_2004_197}. 
Furthermore, the color function constructed throughout this way can also combined with the height-function method \cite{Popinet_2009_228} to enable surface tension calculation in a front-tracking method \cite{Pan_2024_508, Long_2024_513}.

To the authors' knowledge, the only related algorithm for three-dimensional problems was briefly reported by Dijkhuizen et al. \cite{Dijkhuizen_2010_65}. They demonstrated that computing the color function from the geometric information can yield accurate results for certain benchmark cases, such as the stationary bubble and the oscillating drop with viscous damping tests. However, the interface topologies in these test cases are relatively simple, where each Eulerian cell is intersected by the interface at most once. Moreover, for corner cases, in which the element passes exactly through the edge or vertex of the cubic cell, the performance of the segmentation algorithm used to divide triangular elements has not been fully illustrated.

In this work, we propose a more comprehensive algorithm, the Front2VOF (F2V) algorithm, to calculate the volume enclosed by interfaces, discretized into triangular elements, and a cubic cell. 
A robust clipping algorithm is presented to divide elements spanning multiple cubic cells, appropriately accounting for the corner cases.
The F2V algorithm is designed to accurately handle complex topologies, including configurations where a single cubic cell is cut by multiple interfaces. 

The primary objective of the F2V algorithm is to compute the color function for multiphase flow simulations, but its applications extend far beyond this scope.
Given the active research in unstructured mesh generation on curved surfaces described by CAD models, and the availability of well-proven commercial and open-source software, such as \texttt{ICEM-CFD}, \texttt{Gmesh}, for this purpose, the F2V algorithm also provides an accurate method for initializing the volume fraction field in VOF methods.
For interfaces that can be described by analytical functions, the initialization problem has been effectively addressed by various methods, including recursive refinement methods \cite{Cummins_2005_83, Lopez_2009_198} and numerical integration methods, such as those implemented in the \texttt{VOFI} library \cite{Bna_2015_113, Bna_2016_200, Chierici_2022_281}. 
Additionally, Tolle et al. \cite{Tolle_2022_273} proposed an algorithm to calculate the volume fractions from triangulated surfaces immersed in unstructured meshes.
Compared to Tolle's approach, the F2V algorithm leverages the geometric information of elements directly, without relying on an additional level-set function. This capability enables it to handle arbitrarily complex interface topologies with higher computational efficiency.

\section{The Front2VOF algorithm}

To simplify the theoretical derivation and demonstration of the algorithm, we adopt the following assumptions and conventions:

1) Without losing generality, we consider a unit cubic cell with one vertex at the origin: $\Omega_c = \{(x, y, z) ~|~ 0 \leq x, y, z \leq 1\}$. 

2) The six faces of the cube are denoted as $\partial \Omega_{c, x=0, 1} = \{(x, y, z) ~|~ 0 \leq y, z \leq 1, x=0, 1\}$, $\partial \Omega_{c, y = 0, 1}$, and $\partial \Omega_{c, z = 0, 1}$, respectively. 
Cube edges are represented by the intersection of two faces, e.g., $\partial \Omega_{c, x=1} \cap \partial \Omega_{c, z=1}= \{(x, y, z) ~|~ 0 \leq y, \leq 1, x=1, z=1\}$, which denotes the edge shared by the faces $\partial \Omega_{c, x=1}$ and $\partial \Omega_{c, z=1}$.

3) When the cube $\Omega_c$ is intersected by interfaces, it is divided into separated regions occupied by the reference phase, $\Omega_{r, i}, i = 0, 1, 2, ..., N_{r} - 1$, each enclosed by a surface $\partial\Omega_{r, i}$.

In real simulations, such as cases involving jet or thin sheet fragmentation or droplet impact on solid or liquid surfaces, a cube may be divided into complex regions by multiple interfaces.
Regardless of the complexity of the topology of these separated domains, the volume occupied by the reference phase can be computed using Gauss's theorem, relating the flux of a vector field $\FF$ through a closed surface $\partial V$ to the divergence of the field, $\nabla \cdot \FF$, within the volume $V$:
% ----------
\begin{gather}
\iiint_{V} (\nabla \cdot \FF) dV = \iint_{\partial V} (\FF \cdot \hat{\N}) dS,
\label{Eq_Gauss_theorem}
\end{gather}
% ----------
where $\hat{\N}$ is the outward-pointing unit normal vector.

By setting $\FF = (x, 0, 0)$, Eq.~\ref{Eq_Gauss_theorem} simplifies to
% ----------
\begin{gather}
\iiint_{V} dV = \iint_{\partial V} (x \hat{n}_x) dS,
\label{Eq_vol_single}
\end{gather}
% ----------
where the left-hand side of the equation represents the volume of a specific enclosed region.
In the simplest case, where the interface can be described by a single-valued function $x = f(y, z)$, Eq.~\ref{Eq_vol_single} corresponds to the computation of the volume with a 2D integral over the $y$-$z$ plane. 
However, Eq.~\ref{Eq_vol_single} can be applied to more general cases where the interface can not be described by a single-valued function. 
Alternative choices for $\FF$, such as $\FF = (0, y, 0)$ or $\FF = (0, 0, z)$, are equally valid, but the result will not be influenced by the choice of $\FF$ in our algorithm.

By applying Eq.~\ref{Eq_vol_single} to each separated region $\Omega_{r, i}$ and summing the integrals, the total volume of the reference phase inside the cube can be computed as
% ----------
\begin{gather}
V_\Omega = \sum_{i=0}^{N_r - 1}\iint_{\partial \Omega_i} (x \hat{n}_x) dS = \sum_{i=0}^{N_r - 1}\iint_{\partial \Omega_{int, i}} (x \hat{n}_x) dS + \sum_{i=0}^{N_r - 1}\iint_{\partial \Omega_{cube, i}} (x \hat{n}_x) dS,
\label{Eq_vol_multi}
\end{gather}
% ----------
where $N_r$ is the number of separated regions occupied by the reference phase.

% ----------
\begin{figure}
\begin{center}
\begin{tabular}{cc}
\includegraphics[width=0.4\textwidth]{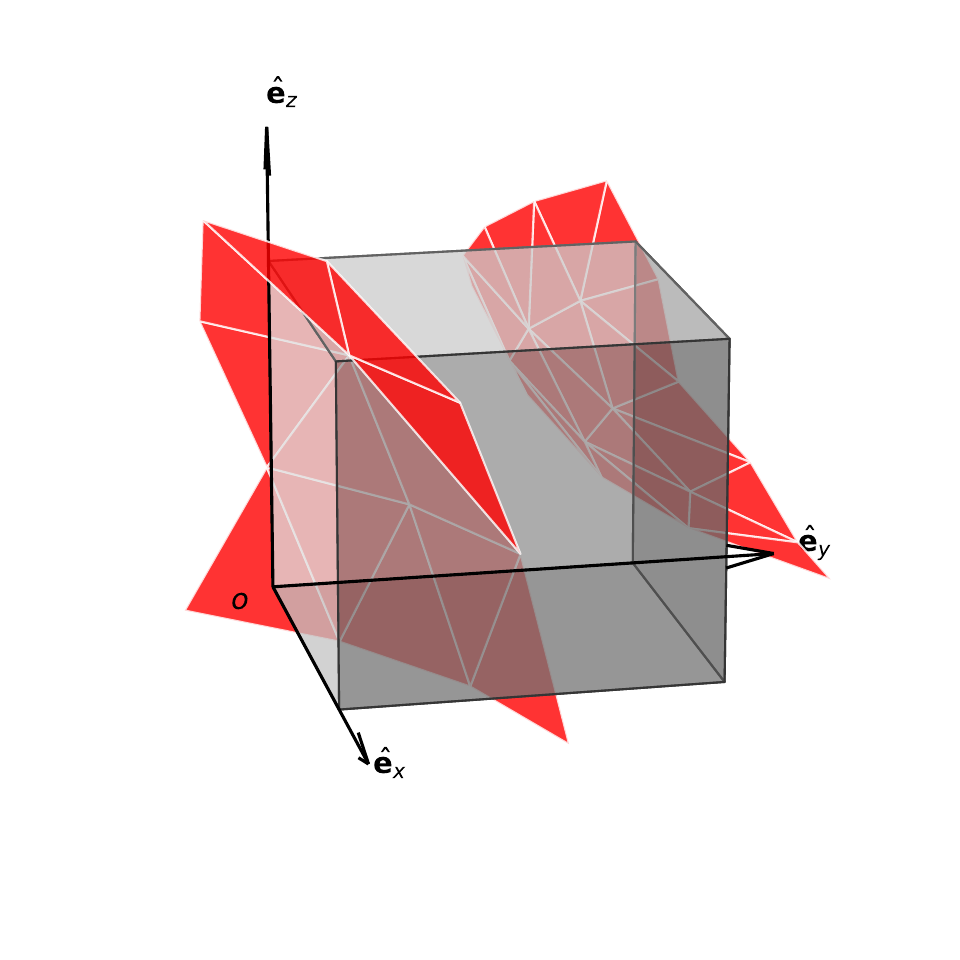} &
\includegraphics[width=0.4\textwidth]{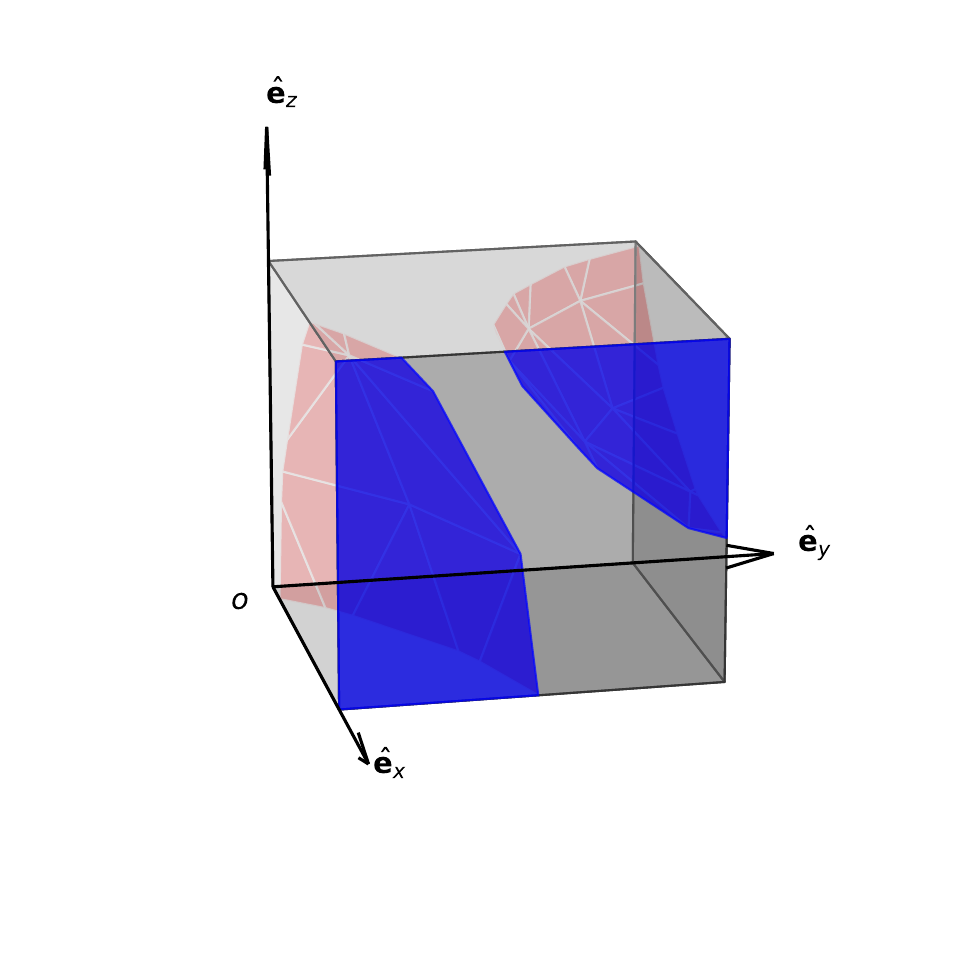}\\
(a)  & (b)  \\
\includegraphics[width=0.3\textwidth]{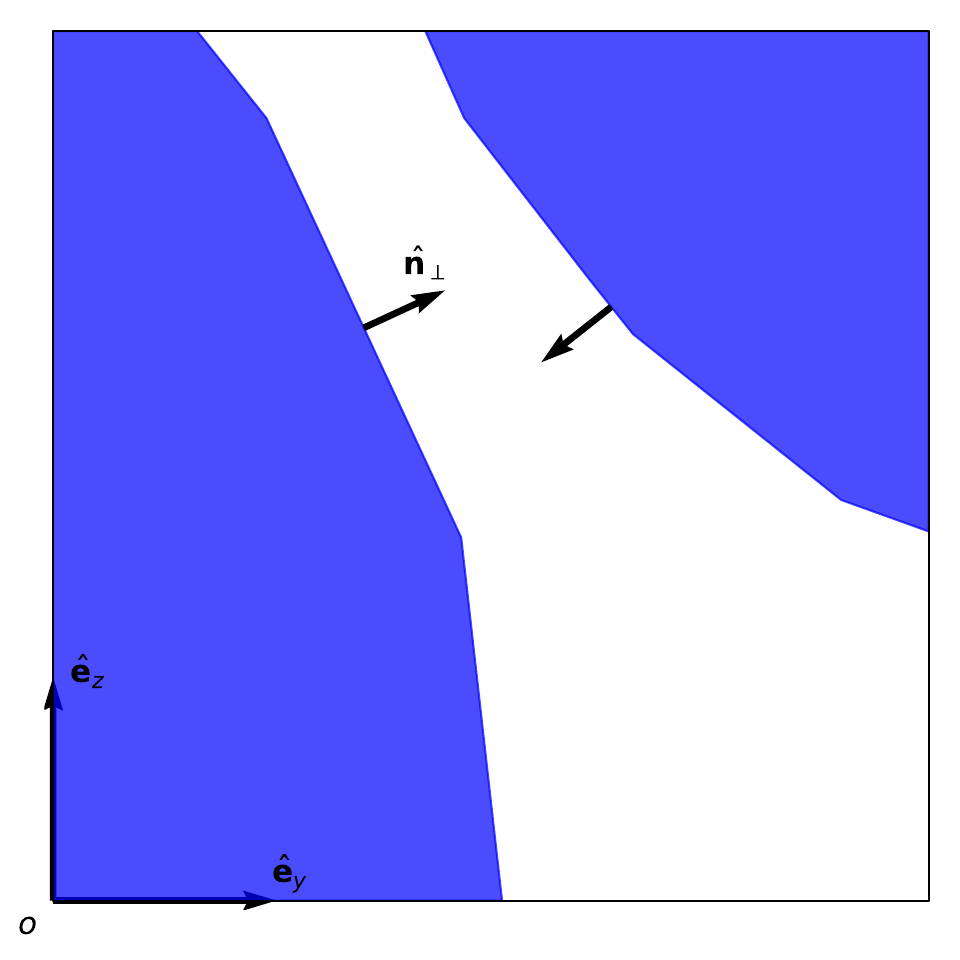}&
\includegraphics[width=0.3\textwidth]{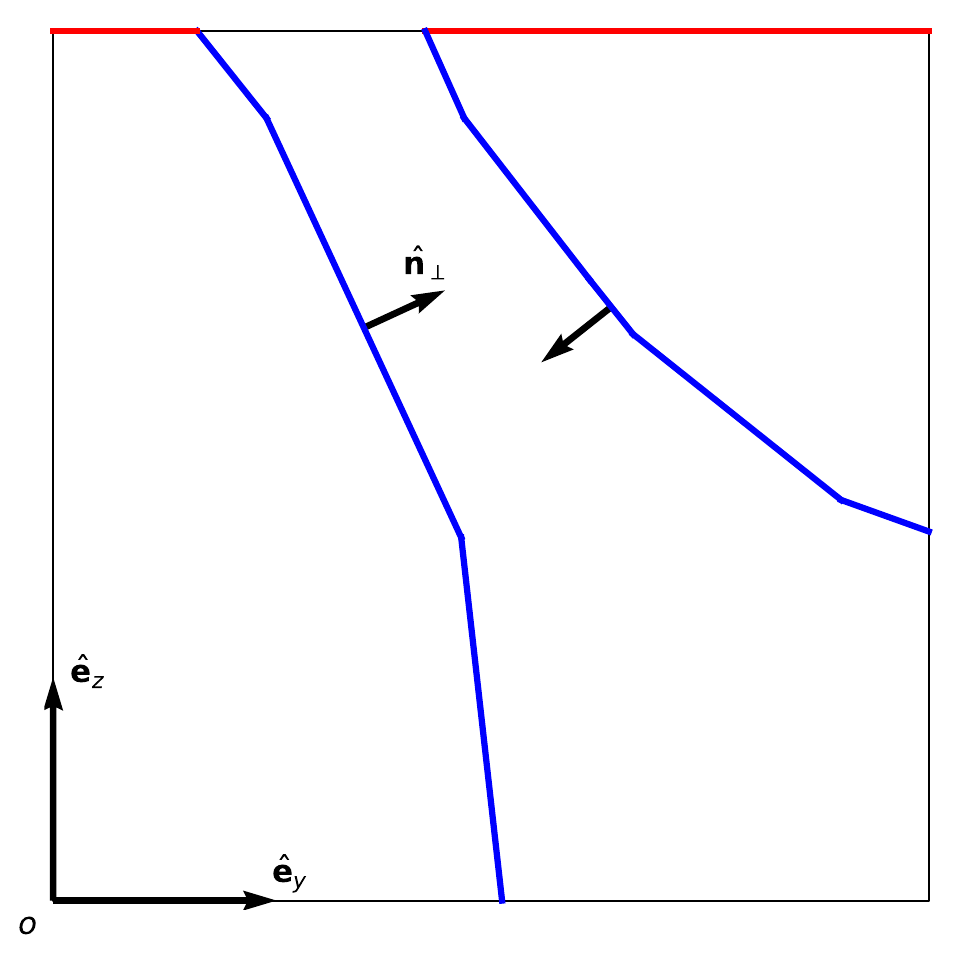}\\
(c)  & (d) \\
\end{tabular}
\end{center}
\caption{Schematic of the Front2VOF algorithm: (a) triangular elements intersecting with a cube; (b) polygons inside the cube obtained using the clipping algorithm; (c) intersection regions between the reference phase and cube face $\partial \Omega{c, x = 1}$; (d) line segments contributing to area computation.}
\label{Fig_f2v_schematic}
\end{figure}
% ----------

The surface integrals in Eq.~\ref{Eq_vol_multi} are further divided into two parts. 
The first summation is contributed by the integrals over the interfaces inside the cube, indicated by the red triangles in Fig.~\ref{Fig_f2v_schematic}b, while the second part represents the integrals over cube faces.

Due to the specific choice of $\FF$ and the convention 1) stated earlier, only the regions on cube face $\partial \Omega_{c, x=1}$, shown as blue areas in Fig.~\ref{Fig_f2v_schematic}b, contribute to the second summation in Eq.~\ref{Eq_vol_multi}. 
Furthermore, since the interfaces are discretized into triangular elements, the surface integrals in the first term can be reformulated as a summation over all interfacial elements within the cube.
In summary, the volume enclosed by the interfaces and the cube can be computed as
% ----------
\begin{gather}
V_\Omega = \sum_{i=0}^{N_{poly} - 1} C_{x, i}\hat{n}_{x,i} S_{i} + S_{\Omega},
\label{Eq_vol_final}
\end{gather}
% ----------
where $N_{poly}$ is the total number of polygons inside the cube, including the origin triangular elements fully contained within the cube, and those formed by intersecting triangles with the cube using the algorithm described in Section \ref{The clipping algorithm}. 
Here, $C_{x, i}$ is $x$-coordinate of each polygon's centroid, $\hat{n}_{x, i}$ is the $x$-component of the unit normal vector, $S_{i}$ is the area of each polygon, and $S_{\Omega}$ is the total area of the intersection regions between the reference phase and the cube face $\partial \Omega_{c, x=1}$.

Similarly, the total area on the cube face, $S_{\Omega}$, can be calculated using Green's theorem, relating the double integral over a plane region $D$ to a line integral around the simple curve $\partial D$ enclosing the region:
% ----------
\begin{gather}
\iint_{D} (\frac{\partial M}{\partial y} - \frac{\partial L}{\partial z}) dS = \oint_{\partial D} (L \hat{t}_y + M \hat{t}_z) dl 
= \oint_{\partial D} (-L \hat{n}_z + M \hat{n}_y) dl,
\label{Eq_Green_theorem}
\end{gather}
% ----------
where $\hat{\T}_{\perp} = (0, t_{\perp, y}, t_{\perp, z})$ is the unit tangential vector along the anticlockwise direction, and $\hat{\N}_\perp = (0, n_{\perp, y}, n_{\perp, z}) = (0, t_{\perp, z}, -t_{\perp, y})$ is the unit normal vector pointing outward from the region $D$.
For an edge of a given polygon, the normal vector on the cube face $\partial \Omega_{c, x = 1}$ is related to the normal vector of the polygon by
% ----------
\begin{equation}
\hat{\N}_\perp = \frac{\hat{\N} - (\hat{\N}\cdot \hat{\E}_x) \hat{\E}_x}{|\hat{\N} - (\hat{\N}\cdot \hat{\E}_x) \hat{\E}_x|},
\end{equation}
% ----------
where $\hat{\E}_x = (1, 0, 0)$ is the unit vector parallel to the $x$-axis.

By setting $L = -z$, $M = 0$, and summing the integrals for all separated regions on the cube face $\partial \Omega{c, x=1}$, we obtain
% ----------
\begin{gather}
S_{\Omega} = \sum_{i=0}^{N_{r} - 1} \iint_{\partial \Omega_{i, x=1}} dS
= \sum_{i=0}^{N_{edge} - 1}(C_{z, i} \hat{n}_{\perp, z, i}) {dl}_i + L_{\Omega}.
\label{Eq_area_final}
\end{gather}
% ----------
where $N_{edge}$ is the total number of polygon edges on the cube face, $C_{z, i}$ is the $z$-coordinate of the edge center, and $dl_i$ is the edge length. 
Similar to the surface integral in Eq.~\ref{Eq_vol_multi}, the line integrals are divided into two parts in the third expression. 
The first part is contributed by the polygon edges on the cube face, denoted by the blue segments in Fig.~\ref{Fig_f2v_schematic}d, while the second part arises from the segments along upper cube edge, $\partial \Omega_{c, x=1} \cap \partial \Omega_{c, z=1}$, shown in red in Fig.~\ref{Fig_f2v_schematic}d.

Employing Eqs.~\ref{Eq_vol_final} and \ref{Eq_area_final} to compute the volume fraction requires identifying only the topology of the intersection regions along a 1D edge explicitly, rather than on a 2D face and in a 3D space, which simplifies the treatment of cases with complex topologies.

It is worth noting that the F2V algorithm can also be generalized for an unstructured tetrahedral mesh with only minor modifications. 
By substituting $\FF$ with $\X - \X_0$, where $\X_0$ represents the coordinates of a selected tetrahedron vertex, Eq.~\ref{Eq_Gauss_theorem} simplifies to
\begin{equation}
3 V_\Omega = \sum_{i=0}^{N_r - 1}\iint_{\partial \Omega_{int, i}} (\X - \X_0) \cdot \hat{\N} dS +  h_{\Omega} S_\Omega,
\end{equation}
where $h_{\Omega}$ denotes the distance from the chosen vertex $\X_0$ to the opposite tetrahedron face, and $S_{\Omega}$ is the area of the intersection region between the reference phase and that tetrahedron face.
The derivations from Eqs.~\ref{Eq_Green_theorem} to \ref{Eq_area_final}, originally developed for a cubic cell, can thus be fully retained,
reducing the problem once more to identifying the separated regions along each tetrahedron edge.

In summary, the F2V algorithm consists of two primary steps. The first step involves identifying the polygons within the cube by segmenting the triangular elements on interfaces (shown as red elements in Fig.~\ref{Fig_f2v_schematic}a) into several polygons and retaining only the portions inside the cube (shown as red elements in Fig.~\ref{Fig_f2v_schematic}b), which will be explained in Section~\ref{The clipping algorithm}. 
The second step computes the volume enclosed by these polygons and the cube faces using
Eqs. \ref{Eq_vol_final} and \ref{Eq_area_final}, which will be presented in detail in Section~\ref{Volume Calculation}.

\subsection{Clipping algorithm} \label{The clipping algorithm}

% ----------
\begin{figure}
\begin{center}
\begin{tabular}{cc}
\includegraphics[width=0.5\textwidth]{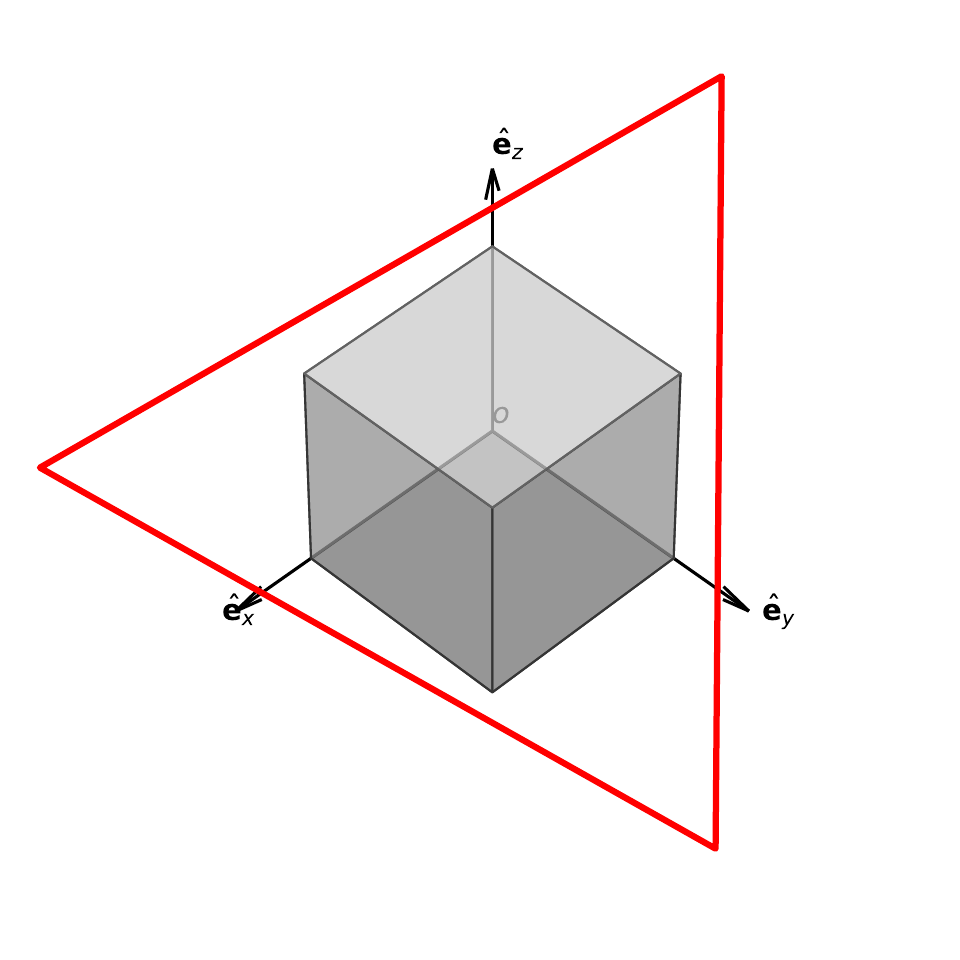} &
\includegraphics[width=0.5\textwidth]{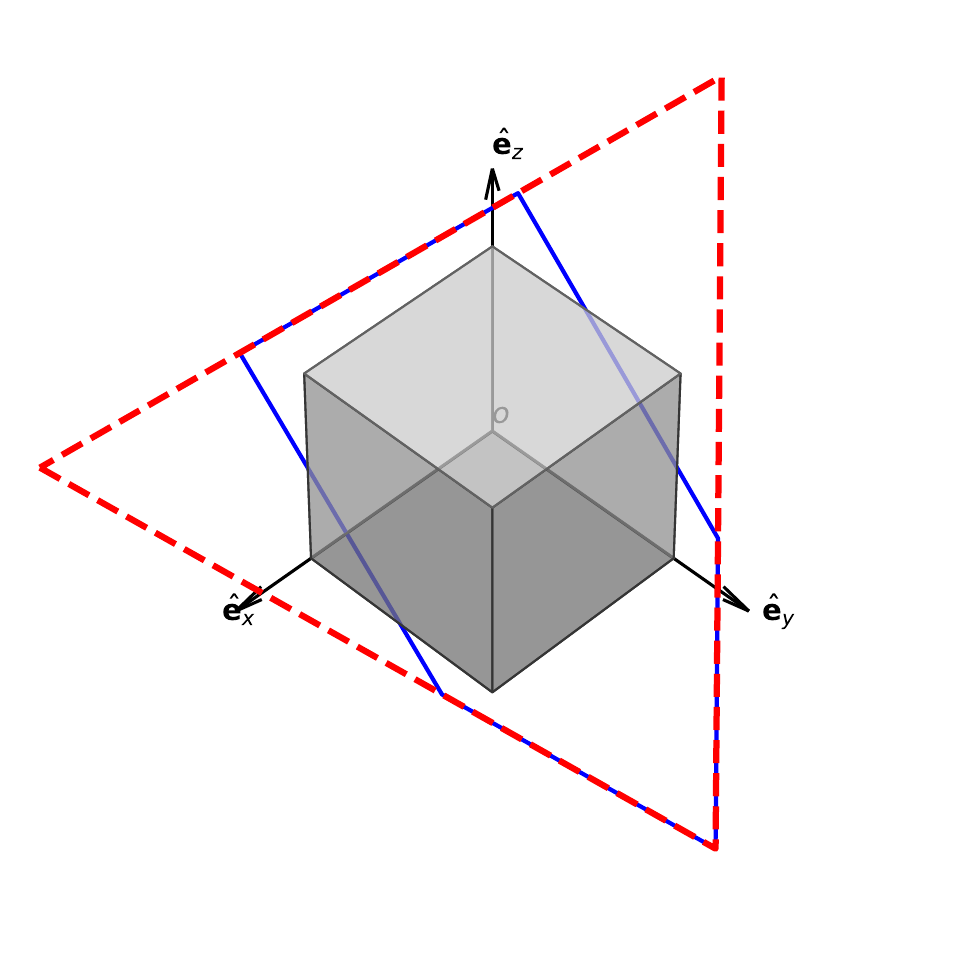}\\
(a)  & (b)  \\
\includegraphics[width=0.5\textwidth]{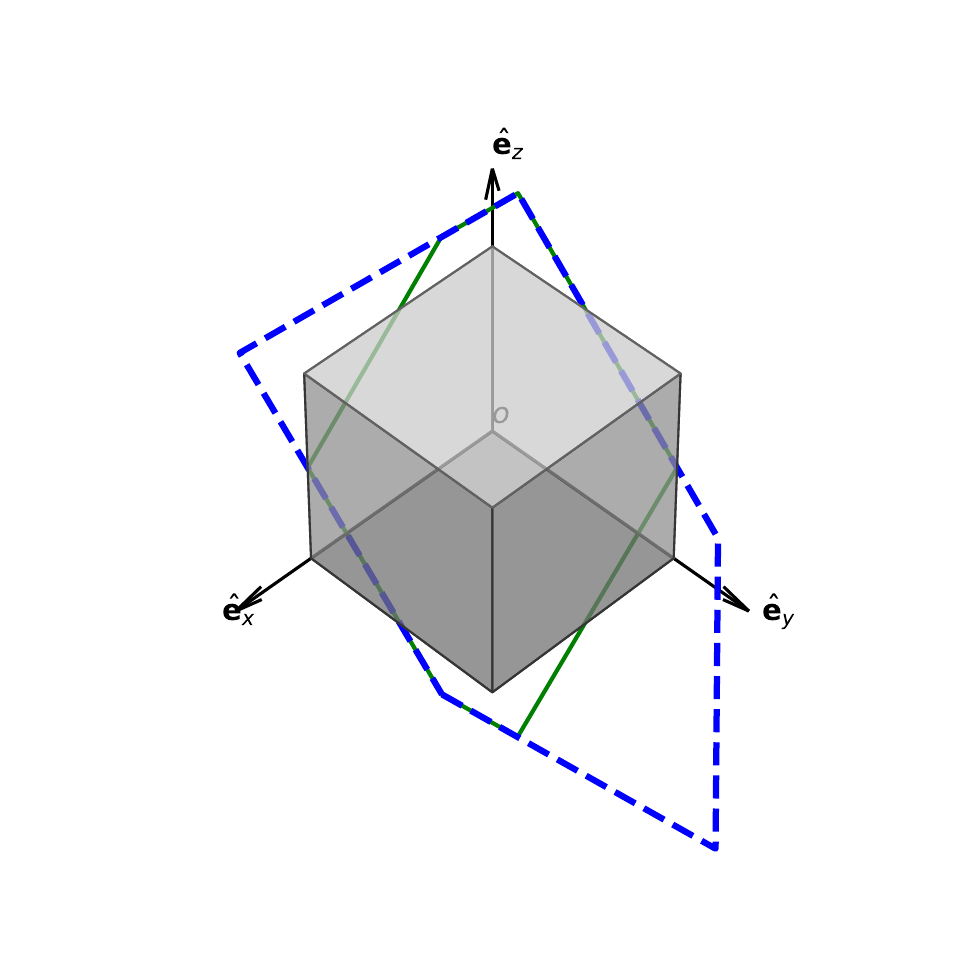}&
\includegraphics[width=0.5\textwidth]{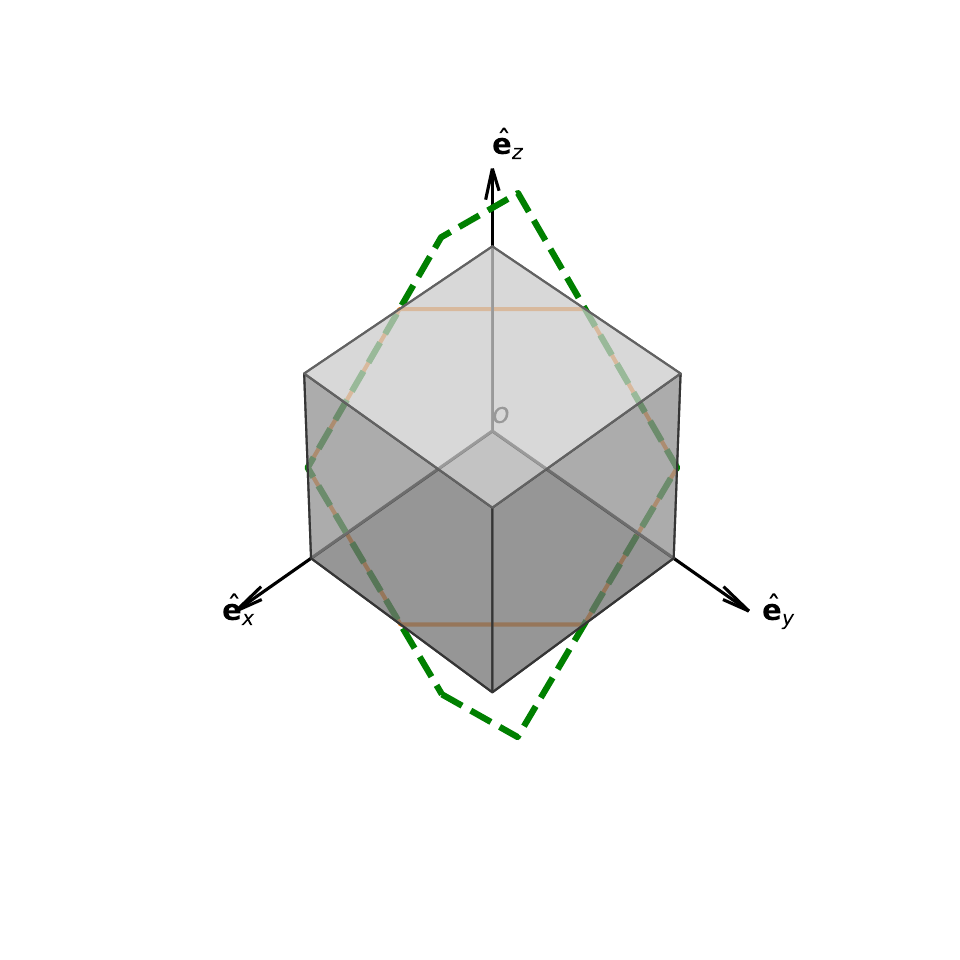}\\
(c)  & (d) \\
\end{tabular}
\end{center}
\caption{Schematic of the clipping algorithm, showing a polygon at different stages of the process: (a) initial triangle, denoted by red solid line; (b) after the clipping by two faces $\partial \Omega_{c, x=0, 1}$, denoted by blue solid line; (c) after the clipping by two faces $\partial \Omega_{c, y=0, 1}$, denoted by green solid line (d) after the clipping by two faces $\partial \Omega_{c, z=0, 1}$, denoted by orange solid line.}
\label{Fig_cut_schematic}
\end{figure}
% ----------

The triangle segmentation procedure is implemented by successively clipping the edges of a polygon with cube faces perpendicular to $x$-, $y$-, and $z$-axes. 
The line clipping algorithm utilized here was originally proposed by Liang and Barsky \cite{Liang_1984_3} and was also employed in the Local Front Reconstruction Method (LFRM) \cite{Shin_2011_230} for interface reconstruction. 
Compared to the approach in Shin's work \cite{Shin_2011_230}, the algorithm presented here is more comprehensive and robust, designed to handle a variety of corner cases effectively.

For a line segment defined by two endpoints $\X_1 = (x_1, y_1, z_1)$ and $\X_2 = (x_2, y_2, z_2)$, the coordinates of the intersection point $\X_{12}$ between the line segment and a plane at $x = x_{p}$ can be computed as:
% ----------
\begin{gather}
\alpha_{12} = \frac{x_{p} - x_1}{x_2 - x_1},\quad
\X_{12} = \X_1 + \alpha_{12} (\X_2 - \X_1).
\label{Eq_line_plane}
\end{gather}
% ----------
The intersection point exists only when:
% ----------
\begin{equation}
x_2 - x_1 \neq 0 \quad \text{and} \quad 0 < \alpha_{12} < 1.
\label{Eq_with_intersect}
\end{equation}
% ----------

First, each edge of the polygon (initially a triangle) is clipped by the two cube faces $\partial \Omega_{c, x=0, 1}$ using Eq.~\ref{Eq_line_plane}. The new intersection points are added to the respective edges, then the polygon vertices located outside the region bounded by the two cube faces are removed. 
This clipping procedure is then repeated successively with the four remaining cube faces in the $y$- and $z$-directions, resulting in the final clipped polygon.

To facilitate the illustration of the segmentation algorithm, we adopt the following conventions:

1) Let $\poly{}{}$ denote a polygon comprising a sequence of vertices $v_i$:
% ----------
\begin{equation}
\poly{}{} = \{v_i ~|~ i = 0, 1, 2, ..., N_{v} - 1 \},
\end{equation}
% ----------
with vertices ordered counter-clockwise relative to the unit normal vector $\hat{\N}$ pointing outward from the reference phase. This convention is commonly adopted in front-tracking codes. A triangle, a subset of the polygon, is denoted by $\tria{}{}$.

2) The coordinates of each vertex are stored in an array $v_i.x[3]$, whose three elements represent the $x$-, $y$-, and $z$-components of the position vector $\X$, respectively.

3) The member $v_i.\text{rm}$ is an integer flag used to identify whether a vertex should be removed after clipping along a given direction. Specifically, $v_i.\text{rm} = 1$ marks a vertex for removal.

4) The cube faces are denoted as face $0, 1, 2, ..., 5$, corresponding to the six cube faces $\partial \Omega_{c, x = 0}$, $\partial \Omega_{c, x = 1}$, $\partial \Omega_{c, y = 0}$, ..., $\partial \Omega_{c, z = 1}$, respectively. Each face is assigned with a unique integer flag $I_{f, i} = 2^i$, where $i$ is the index of the face.

5) For volume computation using Eqs.~\ref{Eq_vol_final} and \ref{Eq_area_final}, polygon edges on cube faces and polygon vertices on cube edges need to be identified. However, the edge-on-face and vertex-on-edge tests based on the position vector of the polygon vertex are limited by the finite floating-point precision of the computer. Therefore, we introduce an integer flag, $v_i.I$, to indicate the positions of the polygon vertex via bitwise operations: $v_i.I = 0$ signifies a vertex not located on a cube face, i.e., inside or outside the cube. While $v_i.I ~\&~ I_{f, j} = I_{f, j}$ indicates that the vertex $v_i$ is precisely located on cube face $j$.
Upon completion of the clipping step along each direction, polygon vertices are assigned the correct flag through Algorithm \ref{Alg_flag}.

% ----------
\begin{algorithm}
\caption{FLAG: Assigning a flag to vertices of a polygon}\label{Alg_flag}
	\KwData{Polygon $\poly{}{}$}
	\KwResult{Polygon $\poly{}{}$ with correct flag for its vertices}
	$\epsilon \gets 1e-12$ \;
	\For{each vertex $v_i$ in $\poly{}{}$}{
		$v_i.I \gets 0$\;
		\For{$idir \gets 0$ \KwTo $2$}{
			\uIf(){$|v_i.x[idir]| < \epsilon $}{
				$v_i.I \gets v_i.I ~|~ I_{f, 2 * idir}$\;
				$v_i.x[idir] = 0.$\;
			}
			\ElseIf{$|v_i.x[idir] - 1.| < \epsilon $}{
				$v_i.I \gets v_i.I ~|~ I_{f, 2 * idir + 1}$\;
				$v_i.x[idir] = 1.$\;
			}
		}
	}
\end{algorithm}
% ----------

Given a triangle on the interface, $\tria{}{}$, the corresponding polygon inside the cube, $\poly{in}{}$, can be determined using Algorithm~\ref{Alg_clip}, CLIP3D. 
The algorithm returns the original triangle if it is entirely contained within the cube, or an empty set if it lies completely outside the cube.
% ----------
\begin{algorithm}
\caption{CLIP3D: Clipping of a triangle by a cube}\label{Alg_clip}
	\SetKwFunction{Flag}{FLAG}
	\KwData{Triangle $\tria{}{}$}
	\KwResult{Polygon inside the unit cube $\poly{}{in}$}
	Initialization: 
	$\poly{}{in} \gets \tria{}{}$, \Flag{$\poly{}{in} $}\;
		
	\For{$idir \gets 0$ \KwTo $2$}{
		$\poly{}{*} \gets \poly{}{in}$\;

		\For{each vertex $v_j$ in $\poly{}{in}$}{
			\lIf(){$v_j.x[idir] > 1.$ \Or $v_j.x[idir] < 0.$}{
				$v_j.rm \gets 1$
			}
			\lElse{
				$v_j.rm \gets 0$
			}
		}
		
		\For{each edge $\pedge{j}$, with two endpoints $v_{j}$ and $v_{j+1}$, in $\poly{}{*}$}{
			$\T = v_{j+1}.\X - v_{j}.\X$\;
			\If(){$\T[idir] \neq 0$}{
				\tcc{If there are two intersections, they should be inserted in ascending order of $\alpha$.}
				\For{$i \gets 0$ \KwTo $1$}{
					$\alpha \gets (i - v_{j}.x[idir]) / t[idir]$\;
					$v^{*}.\X \gets v_{j}.\X + \alpha \T$\;
					$v^{*}.rm \gets 0$\;
					\If(){$0 < \alpha < 1$}{
						Insert $v^{*}$ into to $\poly{}{in}$ on the corresponding edge\;
					}
				}

			}
		}
		Remove the vertex with $v_i.rm = 1$ from $\poly{}{in}$\;
		\Flag{$\poly{}{in} $}\;
		\lIf(){Number of vertices of $\poly{}{in}$ is smaller than 3}{
			\Return{$\emptyset$}
		}
	}
	\Return{$\poly{}{in}$}\;
\end{algorithm}
% ----------

Note that Eq.~\ref{Eq_with_intersect} contains strict inequalities to ensure that edges with endpoints on the cube face remain undivided by the cube face, which thereby prevents the clipping algorithm from producing degenerated polygons with zero-length edges. Furthermore, edges precisely located on the cube face are preserved after segmentation.

Figure~\ref{Fig_cut_schematic} illustrates a schematic of the clipping procedure. 
The initial triangular element is shown in red, while the polygons resulting from the clipping along $x$-, $y$- and $z$-directions are displayed in blue, green, and orange, respectively.

\subsection{Volume Calculation} \label{Volume Calculation}

Once the polygons inside the cube are identified, the precise computation of the volume of reference phase, $V_\Omega$, based on Eqs.~\ref{Eq_vol_final} and \ref{Eq_area_final} can be efficiently carried out using Algorithm~\ref{Alg_F2V}.

\begin{algorithm}
\caption{F2V: The volume calculation algorithm} \label{Alg_F2V}
	\SetKwFunction{Clip}{CLIP3D}
	\KwData{A list of triangles $\{\tria{i}{}\}$ and a cube $\Omega_c$}
	\KwResult{The volume of the reference phase inside the cube,  $V_{\Omega}$}
	Initialization:
	$V_{\Omega} \gets 0,  S_{\Omega} \gets 0, L_{\Omega} \gets 0$, $I_{e} \gets I_{f, 1} ~|~ I_{f, 5}$\;

	\For{each triangle $\tria{i}{}$ in $\{\tria{i}{}\}$}{
		Get the polygon inside the cube:
		$\poly{i}{in} \gets$ \Clip{$\tria{i}{}$}\;

		\If(){All vertices of $\poly{i}{in}$ are located on $\partial \Omega_{c, x = 1}$ \Or $\poly{i}{in} = \emptyset$}{
			\tcc{Ignore the polygon precisely located on $\partial \Omega_{x=1}$ to avoid double counting.}
			\Continue\;
		}
		
		$V_{\Omega} \gets V_{\Omega} + S(\poly{i}{}) \hat{n}_x(\poly{i}{}) C_x(\poly{i}{})$\;

		\For{each edge $\pedge{j}$, with two endpoints $v_{j}$ and $v_{j+1}$, in $\poly{i}{}$}{
		 $I_{e, j} = v_j.I ~\&~ v_{j+1}.I$\;
			\If(){$I_{e, j} ~\&~ I_{f, 1}= I_{f, 1}$ \And $ I_{e, j} ~\&~ I_{e} \neq I_{e}$}{
				\tcc{We skip the polygon edge on cube edge $\partial \Omega_{c, x=1} \cap \partial \Omega_{c, z=1}$ to avoid double count.}
				$S_{\Omega} \gets S_{\Omega} + L(\pedge{j}) \hat{n}_{\perp, z}(\pedge{j})  C_{z}(\pedge{j})$\;
				\If(){$v_j.I ~\&~ I_{e} = I_{e}$}{
					$L_{\Omega} \gets L_{\Omega} + \text{sign}(\hat{n}_{y}) v_j.x[1]$
				}

					\If(){$v_{j+1}.I ~\&~ I_{e} = I_{e}$}{
					$L_{\Omega} \gets L_{\Omega} + \text{sign}(\hat{n}_{y}) v_{j+1}.x[1]$
				}
				
			}
		}
	}
	$S_{\Omega} \gets S_{\Omega} + F_v(L_{\Omega})$\;
	$V_{\Omega} \gets V_{\Omega} + F_v(S_{\Omega})$\;
	
	\Return{$F_v(V_{\Omega})$}\;
\end{algorithm}
% ----------

The centroid, area, and unit normal vector of a polygon $\poly{}{}$ are computed as follows:
% ----------
\begin{gather}
\X_c(\poly{}{}) = \sum_{i=0}^{N_v - 1} v_i.\X/{N_v},\\
S(\poly{}{}) = \sum_{i=0}^{N_v - 1} S_i
, \quad S_i =   \frac{1}{2} |(v_i.\X - \X_c) \times (v_{i+1}.\X - \X_c)|,\\
\CC(\poly{}{}) = \frac{\sum_{i=0}^{N_v - 1} S_i (v_i.\X + v_{i+1}.\X + \X_c)}{3 S(\poly{}{})},\\
\hat{\N}(\poly{}{}) = \frac{(v_0.\X - \X_c) \times (v_{1}.\X - \X_c) }{|(v_0.\X - \X_c) \times (v_{1}.\X - \X_c)|}.
\end{gather}
% ----------
% two role of fv
Note that, at the end of the Algorithm~\ref{Alg_F2V}, the function
\begin{equation}
F_v (x) = \begin{cases}
      x, & \text{if}\ x \geq 0 \\
      1 + x, & \text{if}\ x < 0
\end{cases},
\end{equation}
is introduced to automatically account for the contribution from the cube face $\Omega_{c, x = 1}$ in Eq.~\ref{Eq_vol_final} when there is no intersection between the interface and this cube face. 
Additionally, the function ensures the correct evaluation of the total length $L_{\Omega}$ when the cube edge is intersected multiple times by interfaces. 
For instance, in the configuration depicted in Fig.~\ref{Fig_f2v_schematic}d, $L_\Omega$ is a negative value in Algorithm~\ref{Alg_F2V}, with its absolute value equal to the length of the line segment between the two red regions. The actual total length of the red segment regions is then obtained through the correction provided by $F_v$.

Algorithm~\ref{Alg_F2V} returns a value of $0$ for a cube that is either completely inside or entirely outside the reference phase. 
To address this limitation, we propose Algorithm~\ref{Alg_F2V_EF} to compute the volume for a fully occupied or empty cell based on the signed distance between the cube vertex and the triangular element.
This approach takes advantage of the encoding of the reference phase position in the normal vector of the interfacial element, $\hat{\N}$,
% ----------
\begin{algorithm}
\caption{F2VEF: The volume calculation algorithm for a full/empty cell} \label{Alg_F2V_EF}
	\KwData{A list of triangles $\{\tria{i}{}\}$ and a full/empty cube $\Omega_c$}
	\KwResult{The volume of the reference phase inside the cube,  $V_{\Omega}$}
	Choose the triangle $\tria{min}{}$ closest to the cube center from $\{\tria{i}{}\}$ .\;
	$\X_{tri} \gets v_{0}.\X$\;
	\For{each vertex $\X_{c, i}$ in cube $\Omega_{c}$}{
		\If(){$|(\X_{tri} - \X_{c, i}) \cdot \hat{\N}(\tria{min}{})| > 1 \times 10^{-12}$}{
			\tcc{Cube vertex not on the element plane}
			$\phi \gets \text{sign}(\X_{tri} - \X_{c, i}) \cdot \hat{\N}(\tria{min}{})$\;
			\Break;
		}
	}

	\lIf(){$\phi > 0$}{
		\Return{$1$}
	}
	\lElse{
		\Return{$0$}
	}
	
\end{algorithm}
% ----------

\section{Numerical results and discussion} \label{Numerical results and discussion}

In this section, we present typical configurations obtained by intersecting a cube with a series of synthetic interfaces to validate the F2V algorithm. 
These configurations are specifically chosen to allow for direct comparison with analytical solutions.

The Python source code for the F2V algorithm, along with scripts used to generate the results discussed in this section, is available on the Github repository \cite{Github_f2v}.

\subsection{A cube cut by a single element}

In this test, the cube is intersected by a single equilateral triangle with a side length of $8$ and a variable centroid, $\CC = (\alpha, \alpha, \alpha), 0 \leq \alpha \leq 1$. 
The triangular element is large enough that its intersection with the cube and the cube faces can form an enclosed region.

The plane on which the triangle is located is described by:
% ----------
\begin{equation}
\M \cdot \X = \alpha,
\end{equation}
% ----------
where $\M = (m_1, m_2, m_3)$ is the normalized normal vector with $\sum_{i=1}^{3} m_i = 1$. 
The volume of the reference phase enclosed by the plane and the cube, $V_{\Omega} (\alpha)$, has been provided in Ref. \cite{Scardovelli_2000_164}, and is used here as the reference solution for validation. 

The F2V code has been tested for three different orientations of the normal vector: 
a) $\M = (1, 0, 0)$, b) $\M = (1/2, 1/2, 0)$, and c) $\M = (1/3, 1/3, 1/3)$. 
Furthermore, two possible normal vector directions $\hat{\N} = \pm \M$ are investigated for each orientation.

% ----------
\begin{figure}
\begin{center}
\begin{tabular}{ccc}
\includegraphics[width=0.3\textwidth]{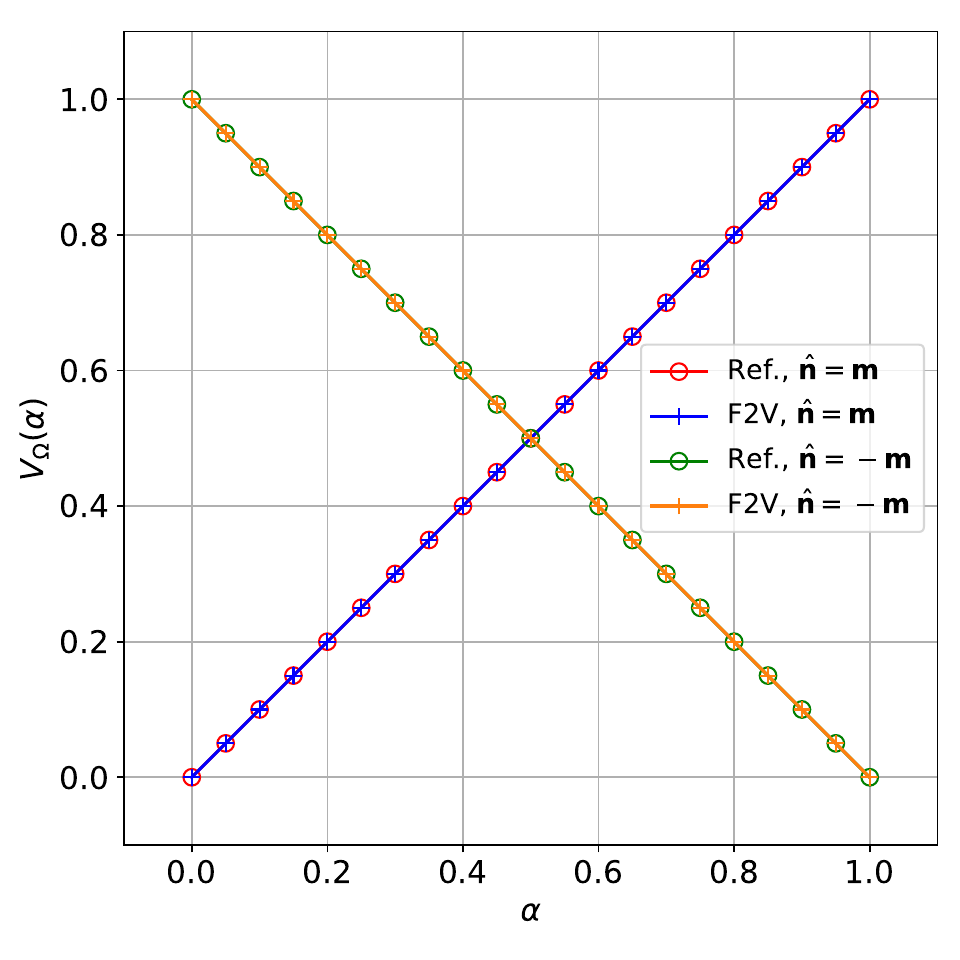} &
\includegraphics[width=0.3\textwidth]{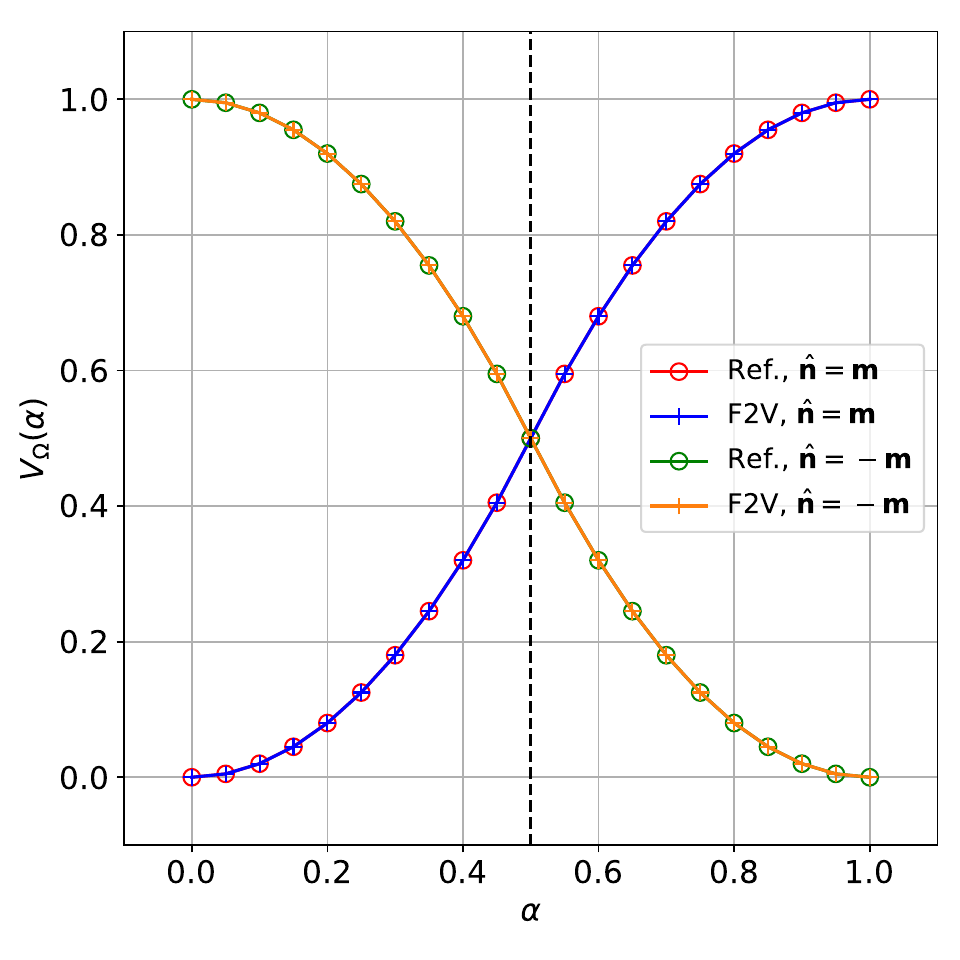} &
\includegraphics[width=0.3\textwidth]{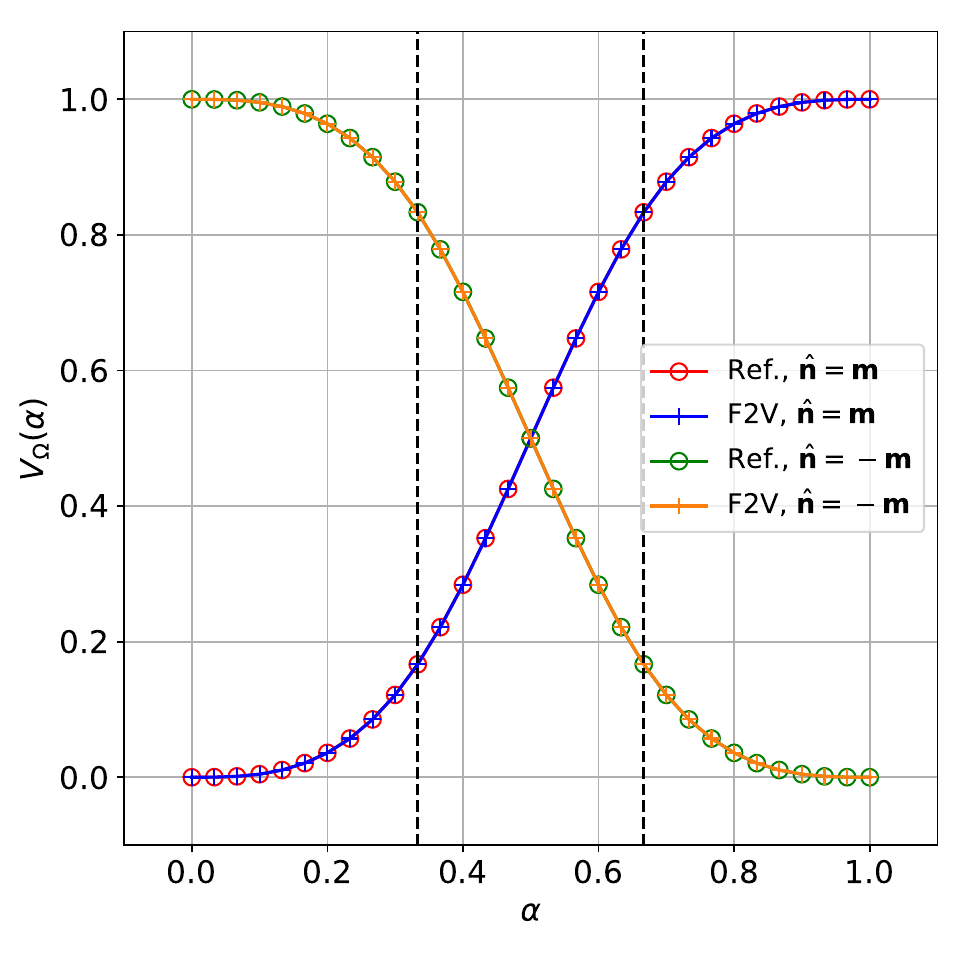}\\
(a) & (b) & (c) \\
\end{tabular}
\end{center}
\caption{A Cube cut by a single element with different normal vectors: $\hat{\N} = \pm \M$,
(a) $\M = (1, 0, 0)$; (b) $\M = (1/2, 1/2, 0)$; (c) $\M = (1/3, 1/3, 1/3)$.}
\label{Fig_volume_single_element}
\end{figure}
% ----------

Figure~\ref{Fig_volume_single_element} shows the variation in computed volume as a function of $\alpha$ for elements with different normal vectors.
The data points obtained using the F2V method coincide precisely with those derived from the analytical formula, with discrepancies between the two methods reaching machine precision.

For $\alpha < 0.5$ in case b or $\alpha < 1/3$ in case c, the element does not intersect with the cube face $\partial \Omega_{c, x = 1}$. 
The results for the opposite unit normal vectors $\hat{\N} = \pm \M$ demonstrate that the contribution from this face is correctly taken into account by the function $F_v(x)$ in Algorithm~\ref{Alg_F2V}. 

Note that specific data points corresponding to corner cases were chosen to highlight the robustness of the triangle clipping method in Algorithm~\ref{Alg_clip}.
For instance, in case $\M = (1/2, 1/2, 0)$, the initial triangular element passes through two opposite edges of the cube when $\alpha = 0.5$.
Similarly, in case $\M = (1/3, 1/3, 1/3)$, the initial triangular element passes through three vertices of the cube when $\alpha = 1/3$ and $\alpha = 2/3$. 
Moreover, for the cases with $\alpha = 0$ and $\alpha = 1$, representing a fully occupied and empty cube, respectively, the phase within the cube is correctly identified by Algorithm~\ref{Alg_F2V_EF}.

\subsection{A cube cut by multiple elements}

In this test, the cube is intersected by two equilateral triangles, each with a side length of $8$. The centroids of these two triangles vary as:
% ----------
\begin{gather} 
\CC_{1} = (\alpha, \alpha, \alpha),\\
\CC_{2} = (g(\alpha), g(\alpha), g(\alpha)),~ g(\alpha) = 1 -\max(\alpha - 0.1, 0). 
\end{gather}
% ----------
The planes on which the triangular elements are located are defined by
% ----------
\begin{gather}
\M_1 \cdot \X= \alpha, \qquad \M_2 \cdot \X = 1. - \max(\alpha - 0.1, 0),
\end{gather}
% ----------
respectively.

For this test, we consider two parallel elements, $\M_1 = \M_2 = (1/3, 1/3, 1/3)$, with opposite normal vectors, $\hat{\N} = \hat{\N}_1 = -\hat{\N}_2$. The two elements move closer to each other as $\alpha$ is increased. To introduce asymmetry, the $\alpha$ of the second plane is shifted by $0.1$. Therefore, the cube is intersected by both elements only when $\alpha > 0.1$, and the two elements coincide with each other at $\alpha = 0.55$.

% ----------
\begin{figure}
\begin{center}
\includegraphics[width=0.9\textwidth]{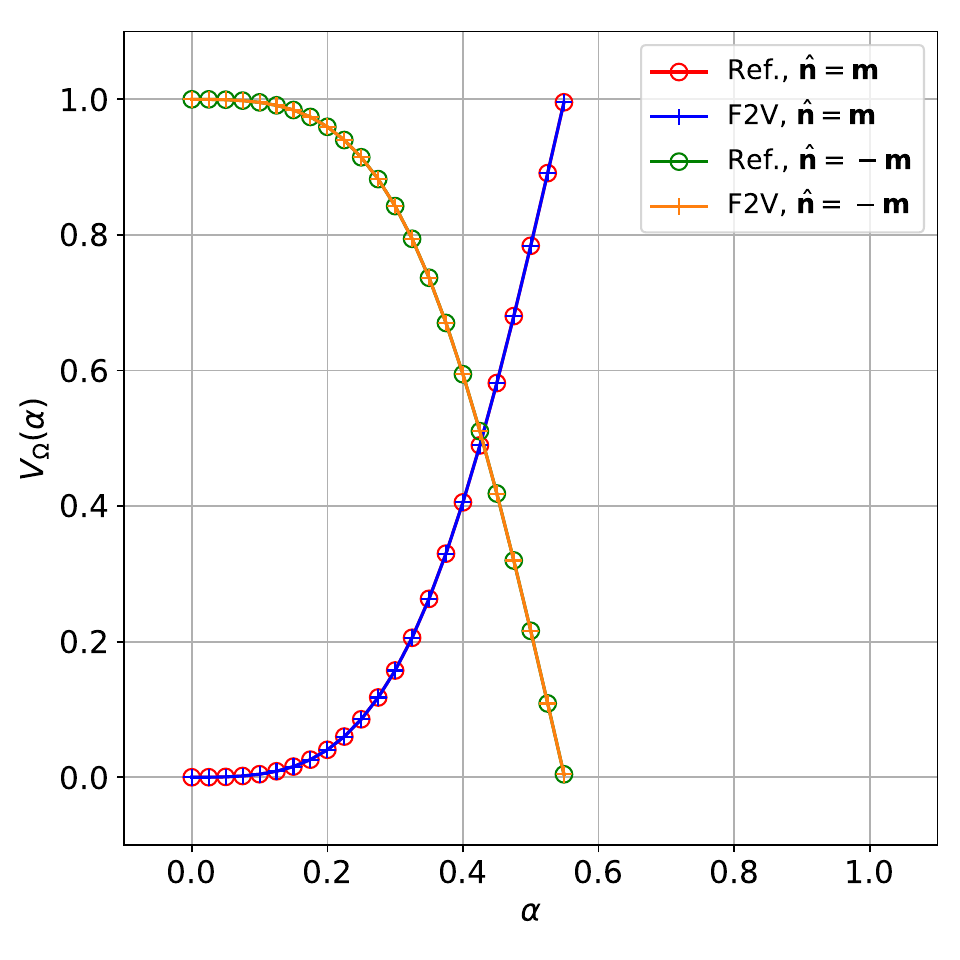}
\end{center}
\caption{A Cube cut by multiple elements: $\hat{\N_1} = -\hat{\N_2}= \pm \M_1$, $\M_1 = \M_2 = (1/3, 1/3, 1/3)$.}
\label{Fig_volume_multiple_elements}
\end{figure}
% ----------

Figure~\ref{Fig_volume_multiple_elements} presents the variation of volume $V_{\Omega}$ as a function of $\alpha$, alongside the reference solutions based on the analytical formula in Ref. \cite{Scardovelli_2000_164}. 
The results obtained with the F2V method agree well with the analytical solutions, with differences remaining at the level of machine precision, as observed in the single-element test.

Notably, when $1/10 < \alpha < 1/3$, the cube is divided into three regions by the two elements, although neither element intersects the cube face $\partial \Omega_{c, x = 1}$. 
As in the previous test, the contribution from this cube face is correctly taken into account by the function $F_v(x)$ when the normal vectors of the two elements are reversed, $\hat{\N} = \pm\M$. 

\subsection{A cube cut by spheres}

In this test, we compute the volume of intersection regions between a cube and multiple spherical interfaces.
This case is more complex than the previous two tests since it includes triangular elements that are entirely outside, entirely inside, or intersecting the cube faces.

Three spherical interfaces with radius $r=0.2$ are centered at different cube vertices: $(0, 0, 0)$, $(1, 1, 1)$, and $(1, 0, 1)$.
The radius is chosen to be small enough to prevent any intersections between these interfaces, though the specific value is arbitrary. 
Placing the spheres at the vertices allows for straightforward computation of the analytical solution, $V_{ana} = \pi r^3 / 2$.
% ----------
\begin{figure}
\begin{center}
\begin{tabular}{ccc}
\includegraphics[width=0.3\textwidth]{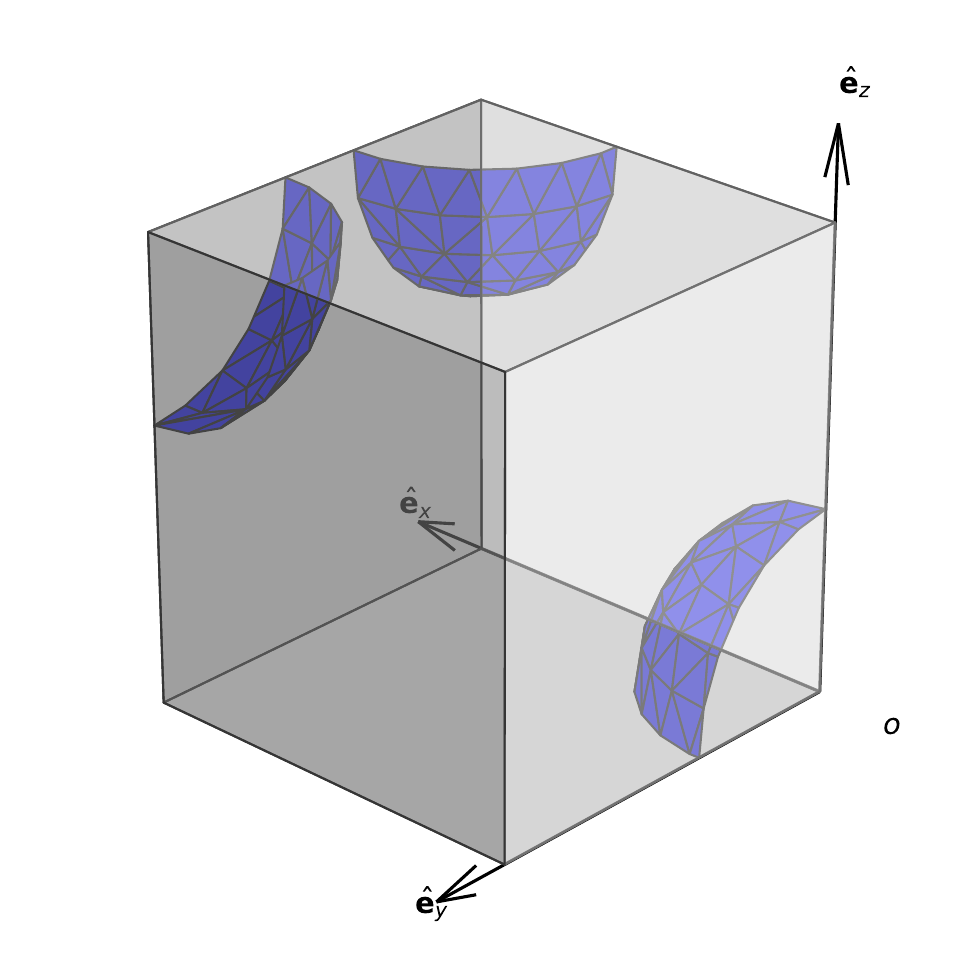} &
\includegraphics[width=0.3\textwidth]{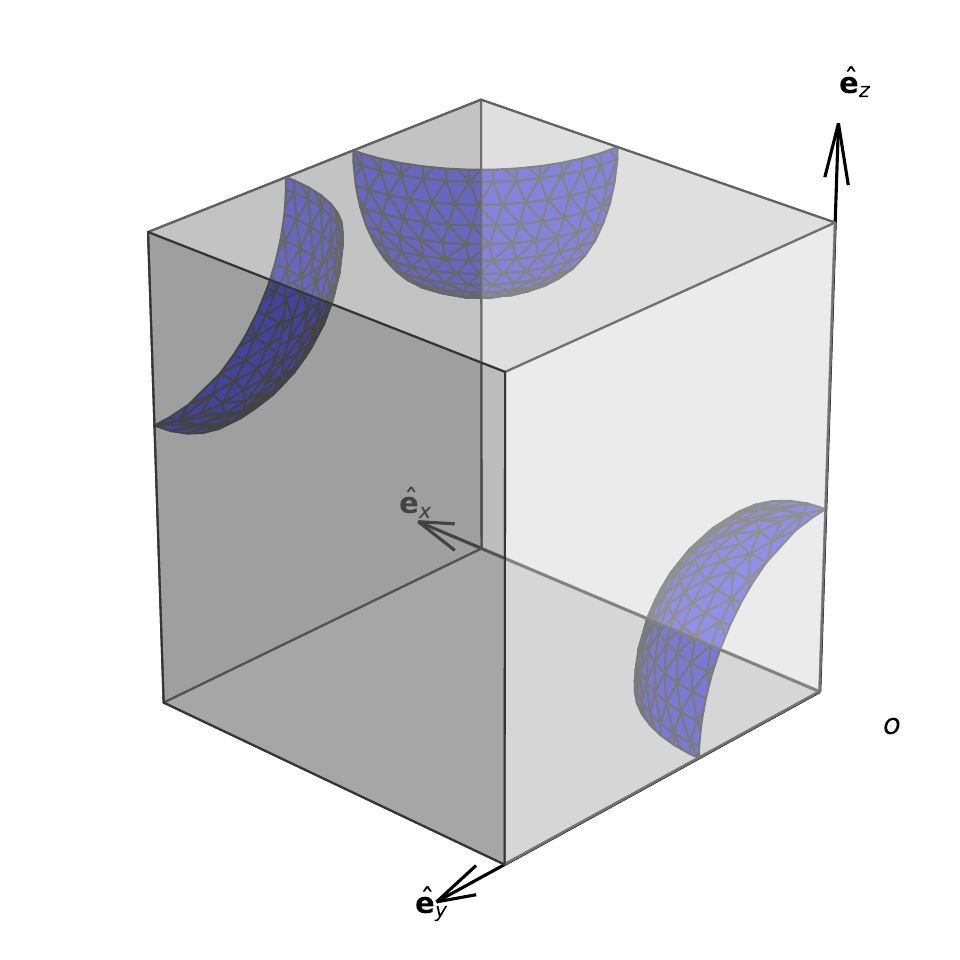}&
\includegraphics[width=0.3\textwidth]{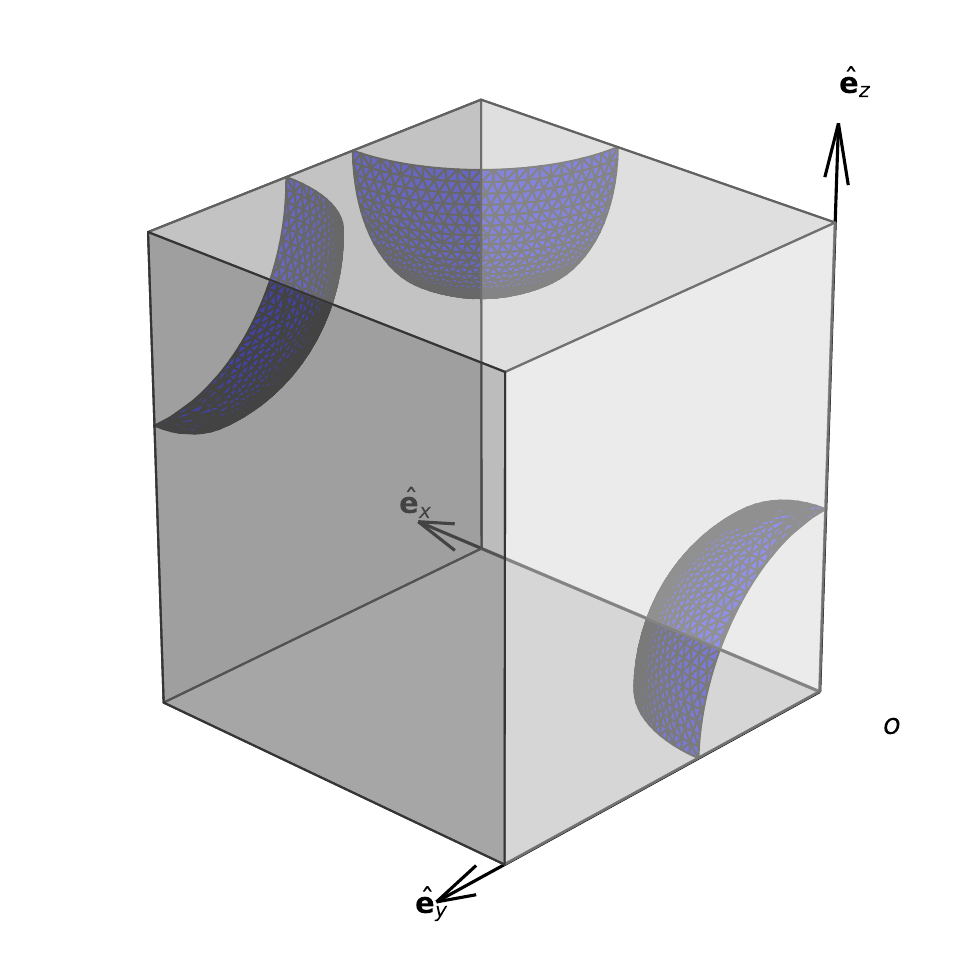}\\
(a) & (b) &(c)\\
\end{tabular}
\end{center}
\caption{A cube cut by spherical interfaces discretized by triangular elements with difference size: (a) $\bar{l} = 0.05$; (b) $\bar{l} = 0.025$; (c) $\bar{l}=0.0125$.}
\label{Fig_volume_sphere}
\end{figure}
% ----------

Only the spherical interface centered at $(0, 0, 0)$ is discretized into triangular elements with a predefined average side length $\bar{l}$. 
The algorithm used to determine the vertices and the connectivity is detailed in Appendix~\ref{appendix-mesh}. 
The meshes on the other two spheres are generated by translating the mesh of the first sphere.

The polygons inside the cube obtained using the clipping algorithm are shown in Fig.~\ref{Fig_volume_sphere} for different element sizes.
Due to the specific positioning of the spheres' center, elements near the pole share a common vertex precisely located on the cube edge, and some edges of the triangular elements lie precisely on the cube face. It has been shown that the clipping algorithm effectively handles these corner-case elements.

The relative volume error, $E_{vol} = |V_{\Omega} - V_{ana}| / V_{ana}$, for different element sizes is listed in Table~\ref{Tab_sphere_con}. 
The results indicate that the volume obtained using the F2V method converges to the analytical value as the element size is decreased, exhibiting an approximately second-order convergence rate.

% ----------
\begin{table}[hbt!]
\caption{Mesh convergence study for a cube cut by spheres.}
\centering
\begin{tabular}{cccc}
\hline 
 $\bar{l}$& $ 0.05$ & $0.025$ & $0.0125$\\ 
\hline 
 $E_{vol}$ & $3.126e-2$ & $7.652e-3$ & $ 2.013e-3$\\ 
\hline 
\end{tabular}
\label{Tab_sphere_con}
\end{table}
% ----------

\section{Conclusions}

We propose a robust algorithm, the Front2VOF (F2V) algorithm, to compute the volume enclosed by a cubic cell and curved interfaces discretized into triangular elements. 
The F2V algorithm consists of two primary steps: 
The first step, termed CLIP3D, identifies the polygons inside the cube by clipping the triangular elements with the cube faces. 
Then the second step computes the volume enclosed by the polygons within the cube and by part of the cube faces.
The F2V method is able to effectively handle the cases where the cube is intersected multiple times by interfaces. 
Moreover, it eliminates the need to reconstruct the complex topology of the enclosed regions on 2D faces and in 3D space. 

To validate the algorithm, we test it using several synthetic test cases of increasing complexity. The good agreement between the computed volumes obtained with the F2V algorithm and the corresponding analytical solutions demonstrates the robustness and accuracy of the algorithm.

\section{Declaration of competing interest}

The authors declare that they have no known competing financial interests or personal relationships that could have appeared to influence the work reported in this paper.

\section{Acknowledgements}

This project has received funding from the European Research Council (ERC) under the European Union's Horizon 2020 research and innovation programme (grant agreement number 883849).

\section{Data availability statement}

The data that support the findings of this study are available from the corresponding author upon reasonable request.

\section{ORCID}

\noindent Jieyun Pan: \url{https://orcid.org/0009-0000-5367-375X}\\
D\'{e}sir-Andr\'{e} Koffi Bi: \url{https://orcid.org/0000-0002-5052-6989}\\
Jiacai Lu: \url{https://orcid.org/0000-0002-0121-9333}\\
Yue Ling: \url{https://orcid.org/0000-0002-0601-0272}\\
Ruben Scardovelli \url{https://orcid.org/0000-0002-1009-2434}\\
Gr\'{e}tar Tryggvason: \url{https://orcid.org/0000-0002-9884-8901}\\
St\'{e}phane Zaleski: \url{https://orcid.org/0000-0003-2004-9090}

\appendix

\section{Mesh generation on a spherical surface} \label{appendix-mesh}

% ----------
\begin{figure}
\begin{center}
\includegraphics[width=0.9\textwidth]{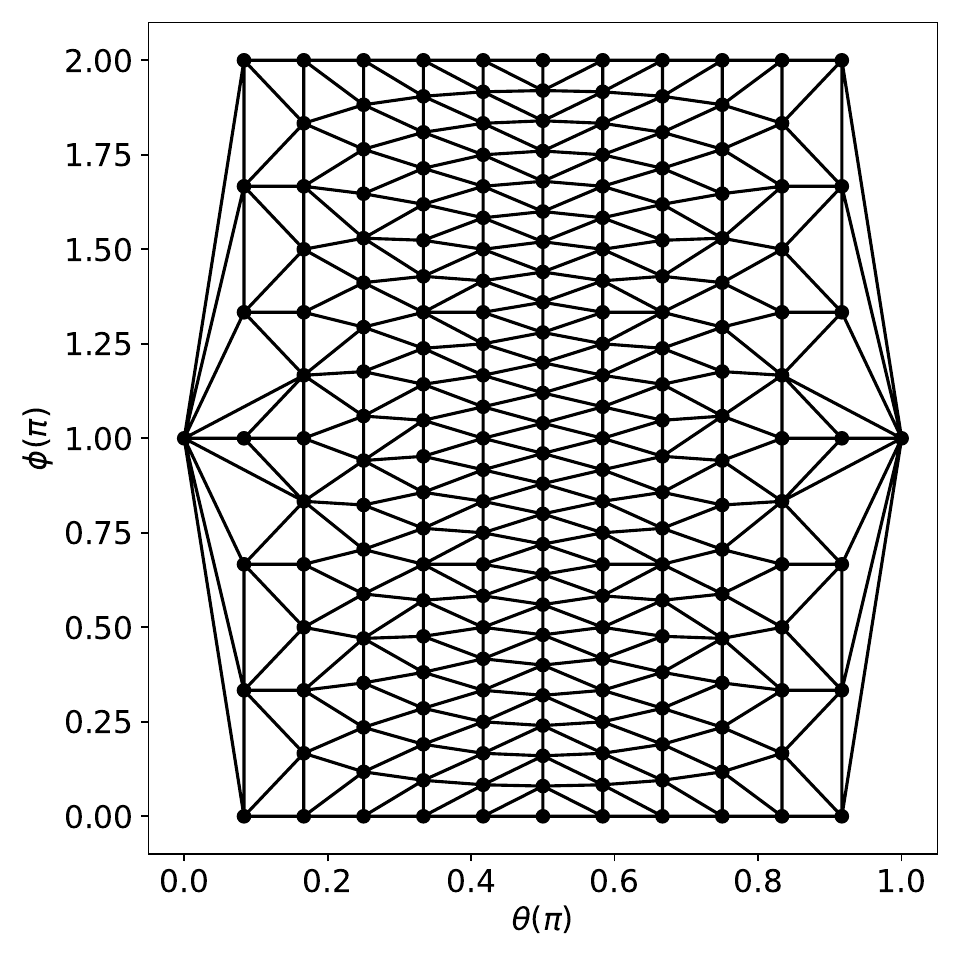}
\end{center}
\caption{Unstructured mesh generation on a spherical surface: vertex distribution and connectivity in $(\theta, \phi)$ space, $\bar{l} = 0.05$.}
\label{Fig_mesh_sphere}
\end{figure}
% ----------

Unstructured mesh generation on a spherical surface is performed using a spherical coordinate system $(r, \theta, \phi)$, where the Cartesian coordinates of a given point are expressed as:
% ----------
\begin{gather}
\label{Eq_spherical_coor}
(x, y, z) = (r \sin\theta \cos\phi, r \sin\theta \sin\phi, r \cos \theta),
\end{gather}
% ----------
with the constraints $r \geq 0$, $0 \leq \theta \leq \pi$ and $0 \leq \pi \leq 2\pi$.

For a fixed radius $r = r_s$, the mesh is generated once the vertex distribution and connectivity in the $(\theta, \phi)$ space have been determined.

Initially, points are distributed in the $(\theta, \phi)$ space based on the parameter $\bar{l}$, which represents the average side length of the triangular elements. 
Two vertices $(\theta, \phi) = (0, \pi)$ and $(\theta, \phi) = (\pi, \pi)$ are first positioned at the poles. 
Then, the $\theta$-direction is uniformly divided into $N_{\theta} = [\pi r_s / \bar{l}]$ intervals, where $[\bullet]$ denotes the floor function.
For each $\theta_i$, the $\phi$-direction is further subdivided into $N_{\phi, i} = [2\pi r_s \sin\theta_{i} / \bar{l}]$ sections.

Once the vertices are determined, the connectivity is established using \texttt{Triangulation} function (class) from the open-source Python library \texttt{matplotlib.tri}, based on Delaunay tessellation. The Cartesian coordinates of the mesh are then computed using Eq.~\ref{Eq_spherical_coor}.
The final vertex distribution on the $(\theta, \phi)$ plane and the connectivity are illustrated in Fig.~\ref{Fig_mesh_sphere} for the case with $\hat{l} = 0.05$.

\clearpage

 \bibliographystyle{model1-num-names}

\bibliography{multiphase-jieyun}
\end{document}